\documentclass{mn2e}
\usepackage{graphicx}
\usepackage{url}
\usepackage{dcolumn}
\newcolumntype{d}{D{.}{.}{2.5}}
\newcolumntype{s}{D{.}{.}{1.2}}
\footnotesize
\newdimen\minuswidth    
\setbox0=\hbox{$-$}
\minuswidth=\wd0
\catcode`@=\active
\def@{\kern\minuswidth}
\newdimen\digitwidth    
\setbox0=\hbox{\rm0}
\digitwidth=\wd0
\catcode`!=\active
\def!{\kern\digitwidth}
\normalsize
\title[Development of a pulsar-based timescale]
{Development of a pulsar-based timescale}
\author[G. Hobbs et al.]
{G. Hobbs,$^1$
W. Coles,$^2$
R. N. Manchester,$^1$
M. J. Keith,$^1$
R. M. Shannon,$^1$
D. Chen,$^3$
\newauthor
M. Bailes,$^4$
N. D. R. Bhat,$^5$
S. Burke-Spolaor,$^{6}$
D. Champion,$^7$
A. Chaudhary,$^1$
\newauthor
A. Hotan,$^1$
J. Khoo,$^1$
J. Kocz,$^4$
Y. Levin,$^8$
S. Oslowski,$^{4,1}$
B. Preisig,$^1$
\newauthor
V. Ravi,$^{9,1}$
J. E. Reynolds,$^{1}$
J. Sarkissian,$^1$
W. van Straten,$^4$
J. P. W. Verbiest,$^7$
\newauthor
D. Yardley,$^{10,1}$
X. P. You$^{11}$
\\
$^1$ CSIRO Astronomy and Space Science, Australia Telescope National Facility, PO~Box~76, Epping
NSW~1710, Australia \\
$^2$ Electrical and Computer Engineering, University of California at San Diego, La Jolla, California, U.S.A. \\
$^3$ National Space Science Center, CAS, Beijing, China \\
$^4$ Centre for Astrophysics and Supercomputing, Swinburne University of Technology, P.O. Box 218, Hawthorn, VIC 3122 \\
$^5$ International Centre for Radio Astronomy Research, Curtin University, Bentley, WA 6102, Australia \\
$^6$ Jet Propulsion Laboratory, 4800 Oak Grove Dr, Pasadena CA 91109-8099, U.S.A. \\
$^7$ Max-Planck-Institut f{\"u}r Radioastronomie, Auf dem H{\"u}gel 69, 53121, Bonn, Germany \\
$^8$ School of Physics, Monash University, P. O. Box 27, Vic 3800, Australia \\
$^9$ School of Physics, University of Melbourne, Vic 3010, Australia \\
$^{10}$ Sydney Institute for Astronomy, School of Physics A29, The University of Sydney, NSW 2006, Australia\\
$^{11}$ School for Physical Science and Technology, Southwest University, 2 Tiansheng Road, Chongqing 400715, China}

\let\leqslant=\leq 
\date{}

\begin{document}
\maketitle
\newcommand{\setthebls}{
}
\setthebls
\begin{abstract}
  Using observations of pulsars from the Parkes Pulsar Timing Array
  (PPTA) project we develop the first pulsar-based timescale that has
  a precision comparable to the uncertainties in international atomic
  timescales.  Our ensemble of pulsars provides an Ensemble Pulsar
  Scale (EPS) analogous to the free atomic timescale \'Echelle
  Atomique Libre (EAL). The EPS can be used to detect fluctuations
  in atomic timescales and therefore can lead to a new realisation of
  Terrestrial Time, TT(PPTA11).  We successfully follow features
  known to affect the frequency of the International Atomic Timescale (TAI)
  and we find marginally significant differences between TT(PPTA11)
  and TT(BIPM11).  We discuss the various phenomena that lead to a
  correlated signal in the pulsar timing residuals and therefore limit
  the stability of the pulsar timescale.
\end{abstract}


\section{Introduction}

Atomic frequency standards and clocks are now the basis of terrestrial
time keeping. Many countries distribute a local atomic
timescale. These are combined by the Bureau International des Poids et
Mesures (BIPM) to form International Atomic Time (or Temps Atomique
International, TAI) which is published in the form of differences from
the national timescales\footnote{The differences between TAI and
  various other timescales can be obtained from the ``Circular T"
  publication available from
  \url{http://www.bipm.org/en/scientific/tai/}.  The difference
  between TT(BIPM) and TT(TAI) is provided at
  \url{ftp://tai.bipm.org/TFG/TT(BIPM)}.}.  TAI is the basis for both
Coordinated Universal Time (UTC), used for the dissemination of time
signals, and Terrestrial Time (TT).  TT is formed by referencing
individual clocks to the Earth's geoid. Throughout this paper, we
refer to TT(TAI) as terrestrial time realised by TAI.  Once published, TAI itself is never revised, but the BIPM publishes another realization of TT which is computed every year and labelled TT(BIPMYY), where YY corresponds to the year of the most recent data used. For instance,
in this paper we refer to TT(BIPM11) as the most recent post-corrected
realisation.

\begin{figure}
\includegraphics[width=6cm,angle=-90]{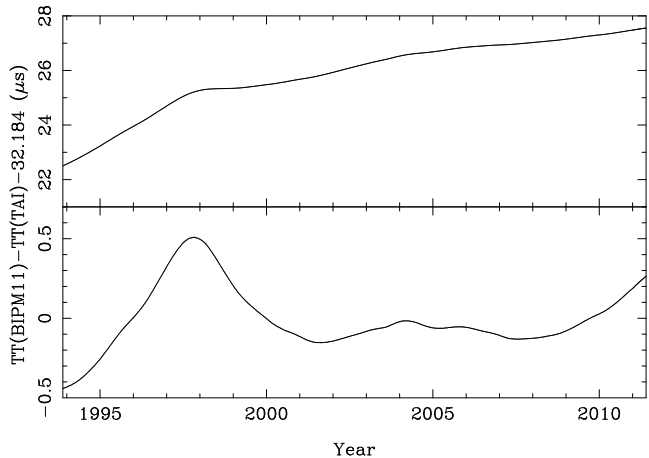}
\caption{The top panel shows the difference between TT(BIPM11) and
  TT(TAI) since the year 1994.  The bottom panel shows the same, but
  after a quadratic polynomial has been fitted and removed.}
\label{fg:tai_bipm}  
\end{figure}

The difference between TT(BIPM11) and TT(TAI) is shown in the top
panel of Figure~\ref{fg:tai_bipm} and clearly shows a drift between
the time standards of $\sim 5$\,$\mu$s since 1994.  The stability of
TAI is obtained from a large number of atomic clocks whereas the
accuracy of TAI is set from a few primary frequency standards (Arias,
Panfilo \& Petit 2011)\nocite{app11}. Initially, the free atomic
timescale \'Echelle Atomique Libre (EAL) is produced from the weighted
average of the timescales of several hundred atomic clocks around the
world.  This timescale is not in accord with the second as defined in
the International System of Units (SI). Therefore, to form TAI (which
does conform to the SI second) from EAL, various frequency adjustments
are necessary.  These are determined using primary frequency
standards.  Frequency adjustments are generally made slowly, a process
referred to as ``steering''.  In 1996, a decision was made to change the realization of the SI second that resulted in a frequency shift of about $2\times10^{?14}$. That shift was progressively introduced into TAI over a period of two years.  As TAI itself is never retroactively corrected, only the
post-corrected versions of TT, e.g., TT(BIPM11), have the earlier data
corrected.  This leads to the ``bump'' that we observe in Figure 1
around the year 1998.

Although numerous clocks are used in forming TAI and there is
continuous development of atomic clocks, stability over decades is
difficult to measure and maintain.  It is therefore desirable to have
an independent precise timescale valid on such long intervals.  In
this paper, we describe the development of such a timescale based on
the rotation of pulsars.

Radio pulsars are rotating, magnetised neutron stars that radiate
beams of electromagnetic waves.  For a fortuitous line of sight to the
pulsar, these can be observed at the Earth as pulses.  The pulse times
of arrival (ToAs) from the brightest and fastest-spinning pulsars can
be measured with a precision of $\sim100$\,ns in an observation time
of $\sim$\,1\,hour.  This precision is significantly worse than that
obtainable from atomic clocks, but, in contrast to individual clocks,
can be maintained for a very long time.  We note that a pulsar-based
timescale provides:
\begin{itemize}
\item{an independent check on terrestrial timescales using a system
    that is not terrestrial in origin.}
\item{a timescale based on macroscopic objects of stellar mass instead
    of being based on atomic clocks that are based on quantum
    processes.}
\item{a timescale that is continuous and will remain valid far longer
    than any clock we can construct.}
\end{itemize}

In order to develop a pulsar-based timescale, all phenomena affecting
the pulse ToAs must be taken into account.  These are incorporated
into a ``pulsar timing model'' that contains the pulsar's astrometric,
rotational and orbital parameters, the effects of the interstellar
medium and the motion of the Earth about the solar system
barycentre. Timing residuals are the difference between the arrival
times converted to the solar system barycentre and predictions of
those times based upon the timing model (see e.g., Edwards, Hobbs \&
Manchester 2006 for details)\nocite{ehm06}.  Non-zero residuals can
result from an incorrect conversion from the measured ToAs to
barycentric arrival times.  Our ability to convert to barycentric
arrival times relies, for instance, upon the accuracy of the solar
system ephemeris.  Many pulsars also display irregularities in
rotation and changes in pulse shape that make timing difficult (e.g.,
Lyne et al. 2011 and references therein\nocite{lhk+10}).  A subset of
pulsars, the ``millisecond pulsars'', have shorter pulse periods and
much more stable rotation than the ``normal pulsars".  However, precise
observations of millisecond pulsars show some unexplained timing
irregularities which we refer to as ``timing noise''.

Some of the variations in the timing residuals are caused by processes
that are correlated between different pulsars.  These can be
identified by observing an ensemble of pulsars, a so-called ``Pulsar
Timing Array'' (PTA, e.g., Foster \& Backer 1990)\nocite{fb90}.  Errors in
the terrestrial time standard will introduce exactly the same signal
in the residuals for each pulsar.  In contrast, errors in the
planetary ephemeris used in the timing analysis will induce timing
residuals which have a dipolar signature on the sky and gravitational
waves propagating past the pulsar and the Earth will induce timing
residuals with a quadrupolar signature. As shown later in this paper,
it is not possible to obtain an unbiased estimate of the time
standard errors simply by forming a weighted average of the timing
residuals for different pulsars. This is because of the coupling
between the timing model for each pulsar and the measurement of the
correlated signal as well as the differing data spans for each pulsar.

In this paper we analyse data from the Parkes Pulsar Timing Array
(PPTA) project (Manchester et al. 2012)\nocite{mhb+12} to develop a
pulsar-based timescale which we label an Ensemble Pulsar Scale (EPS).
This scale has similarities to the free atomic timescale EAL. The
frequency of EAL needs to be steered using primary frequency standards
to realise a timescale based on the SI second.  Similarly, since the
intrinsic pulsar pulse periods and their time derivatives are unknown
for the pulsars in a PTA, the EPS is not an absolute timescale and it
must be ``steered'' to a reference timescale which conforms to the
SI. This is achieved by first forming timing residuals for each pulsar
with respect to the reference timescale, TT(TAI) in our case, and
subsequently fitting a quadratic polynomial to the
residuals. Fluctuations in the reference timescale with respect to the
EPS can be identified and used to provide a set of corrections to that
realisation of TT, thereby realising a new pulsar-based timescale. We
refer to the timescale derived in this paper as TT(PPTA11).  The
bottom panel of Figure~\ref{fg:tai_bipm} shows the difference between
TT(BIPM11) and TT(TAI) after a quadratic polynomial has been fitted
and removed. It is this signal that we expect to see in comparing
TT(PPTA11) with TT(TAI).

Earlier attempts to develop a pulsar timescale have been made by
Guinot \& Petit (1991)\nocite{gp91}, Petit \& Tavella (1996)\nocite{pt96},
Rodin (2008)\nocite{rod08} and Rodin \& Chen
(2011)\nocite{rc11}\footnote{Note that some authors (e.g., Petit \&
  Tavella 1996\nocite{pt96}; Rodin, Kopeikin \& Ilyasov
  1997\nocite{rki97}) have considered using the orbital parameters of
  binary pulsars to provide a pulsar-based timescale.}.  We will show
below that, in contrast to our method, these earlier attempts did not
account fully for the effects of fitting a pulsar timing model. They
also have not been applied to high precision observations for a large
number of pulsars.

In \S2 we describe the signal that is potentially measurable using
pulsar observations.  \S3 describes the observations used in this
paper.  \S4 contains details of the method applied. \S5 presents the
application of our method to actual data and contains a discussion on
the result. \S6 summarises the results.  The algorithm presented here
has been included in the \textsc{tempo2} pulsar-timing software
package (Hobbs, Edwards \& Manchester 2006)\nocite{hem06}.  Usage
instructions are given in Appendix A.

\section{The correlated signal}

Pulsar timing models are based on the proper time, $t^{\rm psr}$, measured at the centre of the pulsar assuming that its gravitational field is not present. Note that the actual time of emission of a pulse and its time of arrival at the solar system barycentre differ by the light travel time from the pulsar, which, for this work, is assumed to be constant\footnote{We note that all pulsars have a radial velocity.  The effect of this velocity is to change the observed pulse frequency by an effectively fixed amount.}. The time of emission of a pulse from the pulsar, $t_e^{\rm psr}$, is therefore related to the observed ToA, $t_a^{\rm obs}$, as
\begin{equation}
t_e^{\rm psr} = t_a^{\rm obs} + \Delta_{\rm clk} + \Delta_{\rm pc} + \Delta_{\rm nc}.
\label{eqn:basic}
\end{equation}
$\Delta_{\rm clk}$ includes all the steps required to convert the measured ToA to barycentric coordinate time (TCB).  The steps (listed below) for this correction are identical for different pulsars.  Any error in this correction will therefore lead to timing residuals for different pulsars that are exactly correlated (i.e., $\Delta_{\rm clk}(t)$ will be identical for all pulsars).  $\Delta_{\rm pc}$ represents steps in the processing that lead to timing residuals that are partially correlated between different pulsars.  For instance, the correlation coefficient may depend upon the angle between the pulsars.  $\Delta_{\rm nc}$ represent corrections that are specific to a given pulsar and are not correlated between different pulsars.  In addition to pulsar dependent effects these corrections include the effects of the interstellar plasma and radiometer noise. 

Errors in the timing system ($\Delta_{\rm clk}$) lead to timing residuals that are correlated between multiple pulsars. $\Delta_{\rm clk}$ can be separated into various components as
\begin{equation}
\Delta_{\rm clk} = \Delta_{\rm trp} + \Delta_{\rm TT} + \Delta_{\rm TCB}.
\end{equation}
$\Delta_{\rm trp}$ represents the time delay between the topocentric reference point of the telescope and the time tagging on the output data.  Most of these delays are constant (for instance, at the Parkes Observatory there is an approximate time delay of 600\,ns from the receiver to the backend instrumentation that records the signal).  However, each backend instrument also has an effective delay and such delays differ for each instrument.  At the Parkes Observatory these delays can be tens of microseconds. A method for measuring and correcting for such delays is described in Manchester et al. (2012).  It is not currently clear how stable these delays are.  Initial studies have suggested that, for some of the backend instrumentation, variations at the 10-100\,ns level may be occurring between observing sessions at the Parkes Observatory.  However, inaccuracies in measuring these delays generally leads to step-changes when a new observing system is commissioned, or adds high-frequency noise if the instrumental time delays randomly change between observations.  These effects are therefore different to the secular drifts expected from errors in terrestrial timescales (see Figure~\ref{fg:tai_bipm}).

$\Delta_{\rm TT}$ is the time difference between the observatory clock and a reference implementation of TT such as TT(TAI).   In order to determine the approximate uncertainty in $\Delta_{\rm TT}$, we compare two independent techniques.  The first method, which is used for our standard data processing, converts from the Observatory time standard to the Global Positioning System (GPS) time standard and uses tabulated corrections from the GPS system to terrestrial time.  The second method uses a GPS Common View system to transfer the Observatory time to the Australian time standard, UTC(AUS).  Tabulated corrections are subsequently used to convert from UTC(AUS) to terrestrial time.  We determined the difference between these two techniques every 10\,d during the year 2011.  The rms difference between the two methods is 8.8\,ns implying that the precision of the time transfer is of this order.   

To obtain barycentric arrival times we have to convert from TT to TCB.  Conversion from TT to TCB is carried out using a time ephemeris described by Irwin \& Fukushima (1999)\nocite{if99}.  In their paper, it is shown that this time ephemeris is known to better than 5\,ns and therefore any errors  will not significantly affect the pulsar timing residuals.     

$\Delta_{\rm pc}$ in Equation~\ref{eqn:basic} represents corrections that, if they are not known with sufficient precision and accuracy, can lead to timing residuals for different pulsars that are partially correlated.  All ToAs are corrected for the geometrical time delay between the Observatory and the solar system barycentre. This is carried out using the Jet Propulsion Laboratory DE421 solar system ephemeris  (Folkner et al. 2008).  Any inaccuracies in this ephemeris will lead to timing residuals whose amplitude depends upon the position of the pulsar with respect to the ecliptic plane.  For two pulsars that are close together on the sky,  ephemeris errors will lead to timing residuals that are correlated.  However, for widely separated pulsars, ephemeris errors will lead to anticorrelated residuals.  At present, errors in the mass of Jupiter and Saturn are the most likely observable effects (Champion et al. 2010); it is very likely that pulsar observations will improve our knowledge of these masses in the next decade.  We will discuss ephemeris errors in more detail in Section~\ref{sec:discussion}. 

The main scientific driver for pulsar timing array observations  is the possibility of detecting gravitational waves. These result in variations in the timing residuals with a quadrupolar signature.  The phenomenon thought to be the most likely to be detected is an isotropic, stochastic, gravitational wave background (Hellings \& Downs 1983)\nocite{hd83}.  Such a background will induce a correlation of $-0.15 < \zeta < 0.5$ between pulsar pairs depending upon the angle between the pulsars.  For our sample of pulsars, the mean $|\zeta|$ is $0.15$, mean $\zeta$ is $0.02$ and the maximum $\zeta$ is $0.42$ for PSRs J1730$-$2304 and J1744$-$1134.  We discuss the possibility that a gravitational wave signal could be misidentified as a clock error in Section~\ref{sec:discussion}.

\section{Observations}

\begin{figure}
\includegraphics[width=8.5cm]{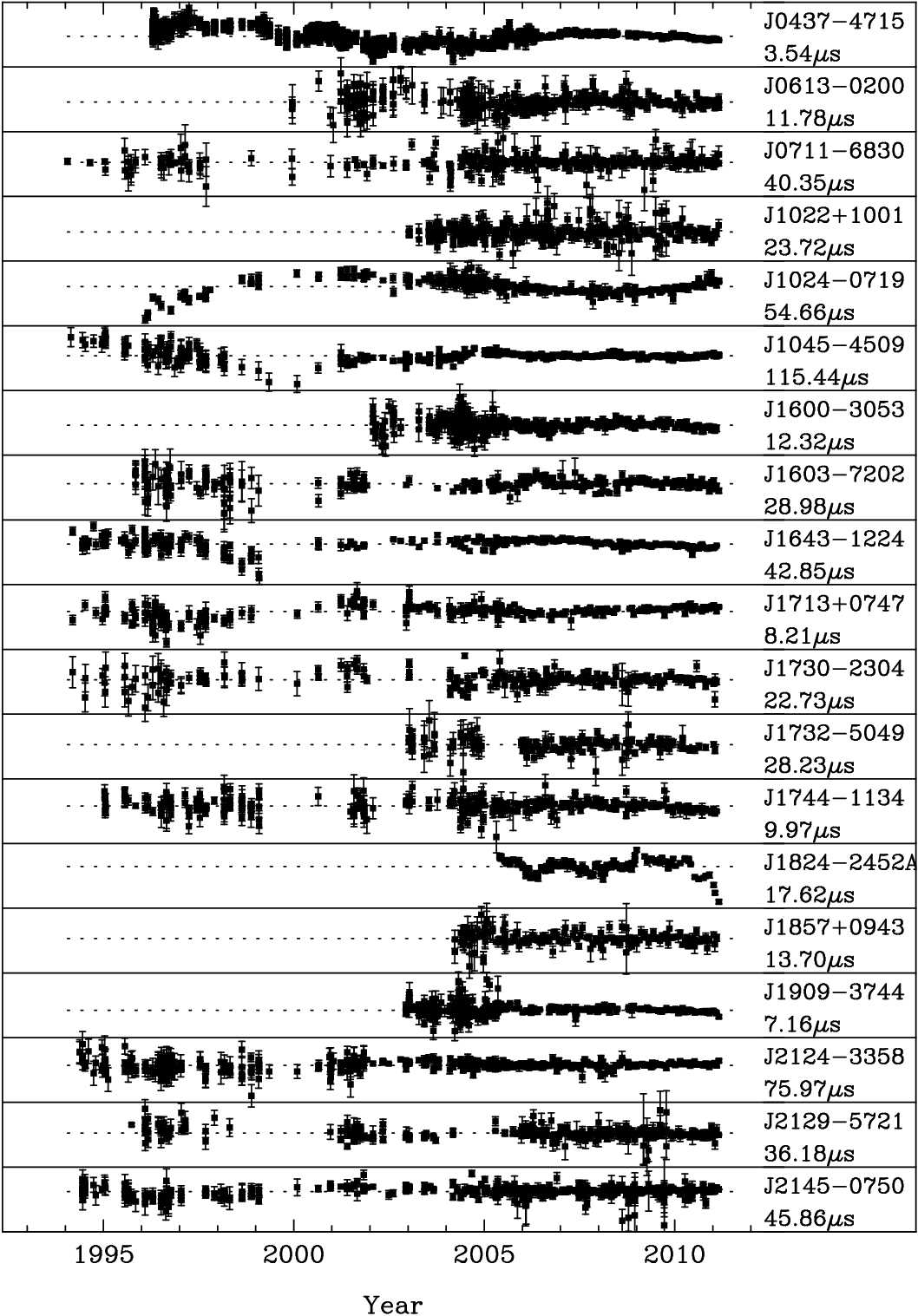}
\caption{The timing residuals used in this analysis referred to TT(TAI).  The label on the right-hand side gives the pulsar name and the total range of the residuals for that pulsar.}
\label{fg:tresiduals}  
\end{figure}

The data used here are the extended Parkes Pulsar Timing Array (PPTA) data set that is described in Appendix A of Manchester et al. (2012).  This data set included observations taken as part of the PPTA project that commenced in the year 2005 along with earlier observations published by Verbiest et al. (2008, 2009)\nocite{vbv+08}\nocite{vbc+09}.  All observations  were obtained using the Parkes 64-m radio telescope.   Typical observation durations were 1\,hr.  The observing system has improved significantly since the earliest observations and so the ToA uncertainties have generally decreased with time.  We have measured the majority of the timing offsets between the different observing systems.  Hence, we have been able to remove most, but not all, of the arbitrary offsets from the timing model that were included in the Verbiest et al. (2008, 2009) analyses (details are given in Manchester et al., 2012).  As the timing residuals for PSR~J1939$+$2134 exhibit timing noise that is currently uncorrectable and is at a level significantly higher than the white noise level, we have not included this pulsar in our analysis.

Observations since 2005 have been corrected for variations in the ionised interstellar medium using multi-frequency observations, but observations prior to 2005 could not be corrected because adequate multi-frequency observations were not made.  Dispersion measure fluctuations are a significant noise source for many of the pulsars in the PPTA (You et al. 2007)\nocite{yhc+07}, so our inability to correct them in the early observations leads to larger error bars on the estimated TT(PPTA11) before 2005.

Timing residuals were formed using the \textsc{tempo2} software using the JPL DE421 solar system ephemeris and referred to TT(TAI)\footnote{This is in contrast to the same data set described by Manchester et al. (2012). In that paper the data were referred to TT(BIPM11).}.  The timing residuals for the 19 pulsars are shown in Figure~\ref{fg:tresiduals} and a summary of our data sets is provided in Table~\ref{tb:dsets} where, in column order, we provide: 1) the pulsar name, 2) pulse period, 3) dispersion measure, 4) weighted rms residual, 5) unweighted rms residual, 6) median ToA uncertainty, 7) data span, 8) number of observations, 9) date of first observation and 10) date of most recent observation.  We emphasise that the following properties of the data set must be accounted for in the analysis:
\begin{itemize}
\item{Each pulsar has a different data span.  For some pulsars data exist from 1995 onwards.  For other pulsars only $6-8$\,yr of data exist.}
\item{The data sampling is irregular with the more recent data being more uniform than earlier data.}
\item{Very few observations were made around the year 2000.  Only PSR~J0437$-$4715 provides significant data around this time.}
\item{The ToA uncertainties are variable.  They generally decrease with time as new instruments were commissioned.  However, pulsar scintillation also leads to significant variations in the uncertainties. It is also common that the uncertainties underestimate the white noise present in the data.  We account for this by including scaling factors that increase the error bars.  In Table~\ref{fg:tresiduals} the median ToA uncertainty is determined without these extra scaling factors.}
\item{Timing noise is observed in many of the data sets.}
\item{The rms timing residuals vary widely. The smallest weighted rms residual is 0.23\,$\mu$s for PSR~J0437$-$4715, whereas the largest is 5.1\,$\mu$s for PSR~J1045$-$4509.  Because of red noise in many of the data sets,  these rms values are often larger than the typical ToA uncertainties.} 
\end{itemize}

\begin{table*}
\caption{Parameters for the pulsar timing residuals referred to TT(TAI).}\label{tb:dsets}
\begin{tabular}{lddddddslll}
\hline
PSR J & \multicolumn{1}{c}{Period} &  \multicolumn{1}{c}{DM} &  \multicolumn{1}{c}{Weighted} &  \multicolumn{1}{c}{Unweighted} &  \multicolumn{1}{c}{Median ToA} & \multicolumn{1}{c}{Span} &  \multicolumn{1}{c}{N$_{\rm obs}$} &  \multicolumn{1}{c}{First} &  \multicolumn{1}{c}{Last} \\
             &         & &  \multicolumn{1}{c}{rms}  &  \multicolumn{1}{c}{rms} &  \multicolumn{1}{c}{uncertainty} &  \\
             &  \multicolumn{1}{c}{(ms)} &  \multicolumn{1}{c}{(cm$^{-3}$pc)}  &  \multicolumn{1}{c}{($\mu$s)} &  \multicolumn{1}{c}{($\mu$s)}  &  \multicolumn{1}{c}{($\mu$s)} &  \multicolumn{1}{c}{(yr)} & &  \multicolumn{1}{c}{(MJD)} & \multicolumn{1}{c}{(MJD)}\\ 
\hline
J0437$-$4715 & 5.757 & 2.65    & 0.23 & 0.36 & 0.23 & 14.8 & 3322 & 50191 & 55618 \\
J0613$-$0200 & 3.062 & 38.78  & 1.12 & 1.71 & 0.92 &11.2 & 281 & 51527 & 55618 \\
J0711$-$6830 & 5.491 & 18.41  & 1.53 & 5.02 & 2.34 &17.1 & 319 & 49374 & 55619 \\
J1022$+$1001 & 16.453 & 10.25 & 2.38 & 3.26 & 1.23 & 8.1 & 378 & 52650 & 55618 \\
J1024$-$0719 & 5.162 & 6.49  & 4.35 & 6.66 & 2.96 &15.1 & 309 & 50118 & 55620 \\ \\

J1045$-$4509 & 7.474 & 58.15 & 5.05 & 11.28 & 3.50 &17.0 & 393 & 49406 & 55620 \\
J1600$-$3053 & 3.598 & 52.19  & 0.99 & 1.45 & 0.60 &9.0 & 503 & 52302 & 55598 \\
J1603$-$7202 & 14.842 & 38.05  & 2.18 & 3.88 & 1.41 & 15.3 & 290 & 50026 & 55618\\
J1643$-$1224 & 4.622 & 62.41  & 2.06 & 4.84 & 1.32 & 16.9 & 288 & 49422 & 55598 \\
J1713$+$0747 & 4.570 & 15.99  & 0.46 & 0.92 & 0.40 & 17.0 & 318 & 49421 & 55619\\ \\

J1730$-$2304 & 8.123 & 9.61  & 2.61 & 3.20 & 1.46 & 16.9 & 223 & 49422 & 55598 \\
J1732$-$5049 & 5.313 & 56.84 & 2.49 & 3.81 & 2.48 & 8.0 & 149 & 52647 & 55581 \\
J1744$-$1134 & 4.075 & 3.14 & 0.67 & 1.28 & 0.54 & 16.1 & 368 & 49729 & 55598 \\
J1824$-$2452A & 3.054 & 119.86  & 2.07 & 2.18 & 0.49 & 5.7 & 178 & 53519 & 55619 \\
J1857$+$0943 & 5.362 & 13.31 & 0.96 & 1.87 & 1.21 & 6.9 & 152 & 53087 & 55598 \\ \\

J1909$-$3744 & 2.947 & 10.39 & 0.20 & 0.60 & 0.22 & 8.2 & 693 & 52618 & 55618 \\
J2124$-$3358 & 4.931 & 4.62 & 2.92 & 7.30 & 2.43 & 16.8 & 473 & 49490 & 55618 \\
J2129$-$5721 & 3.726 & 31.85 &  1.40 & 3.90 & 2.32 & 15.4 & 285 & 49987 & 55618\\
J2145$-$0750 & 16.052 & 9.00 &  1.05 & 3.89 & 1.48 & 16.7 & 696 & 49517 & 55618\\
\hline
\end{tabular}
\end{table*}

\section{Method}\label{sec:method}

\begin{figure}
\includegraphics[width=6cm,angle=-90]{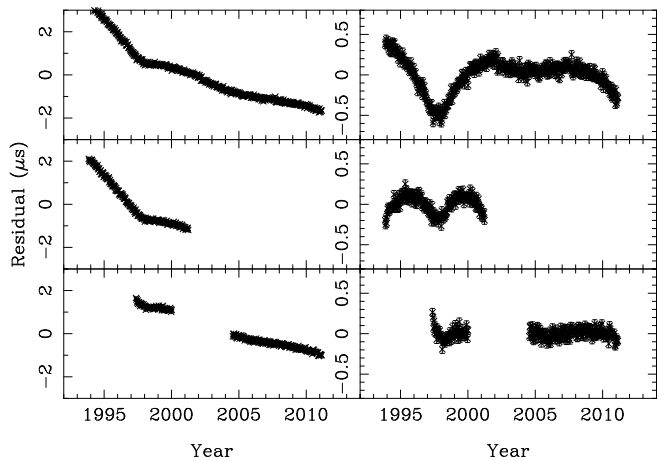}
\caption{Simulated timing residuals for three pulsars in the presence of the deliberate steering of TAI.  The left-hand panel shows the pre-fit residuals for the three pulsars.  The right-hand panel (which has a different y-scaling) shows the post-fit residuals.}\label{fg:sim1}
\end{figure}
\begin{figure}
\includegraphics[width=6cm,angle=-90]{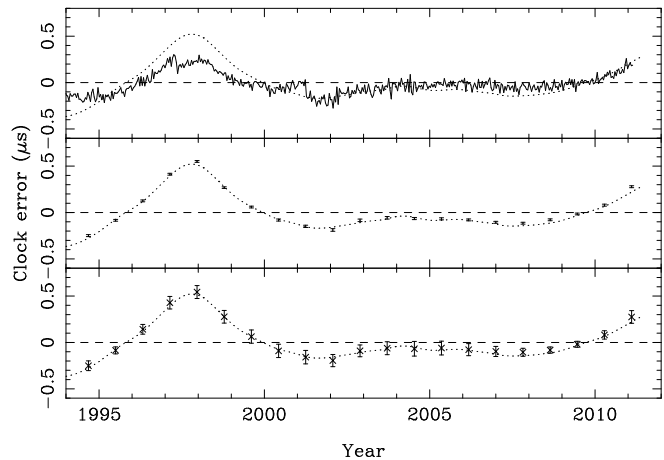}
\caption{In all panels the dotted line represents TT(BIPM11)-TT(TAI) with a quadratic polynomial fitted and removed.   The solid line in the top panel shows the result from a simple weighted average of the simulated timing residuals that are shown in Figure~\ref{fg:sim1}. The data points in the central and lower panels are the result of the algorithms described in this paper.  In the lower panel the error bars account for the possibility of timing noise in the timing residuals for each pulsar.}\label{fg:compareWeight}
\end{figure}

To include clock errors in the timing model we need a function that describes the clock error, $\Delta_c (t)$, at any time, $t$, during the data span, in terms of a small number of parameters. We want to avoid imposing any more structure on this model than necessary. We tried two approaches: a Fourier series and a set of equally spaced samples with an interpolation mechanism. Both provided adequate results, but we found the set of equally spaced samples provided better error estimates and more flexibility with the model parameters.  We use linear interpolation between the samples which have spacing $T_s$; this is equivalent to a low-pass filter with a bandwidth of $f_{LP} = 1/2T_s$. Regardless of the functional form of the model, the parameterisation of the clock error will have some covariance with the other parameters in the timing model. We implemented constraints on the clock parameters to minimise this covariance.  These constraints zero the offset, linear and quadratic terms in $\Delta_c(t)$, and those terms that represent a position error, parallax or proper motion.  The changes to the least-squares-fitting algorithm that enable these constraints are described in detail by Keith et al. (2012)\nocite{kcs+12}.  As these terms are removed from the individual pulsar residuals they must not exist in the clock error because they would be unconstrained.  The modifications to the standard \textsc{tempo2} least-squares-fitting procedure to allow the $\Delta_c(t)$ values to be fit globally to all these data sets were originally described by Champion et al. (2010)\nocite{chm+10}.

It would be simpler to form the weighted average of the timing residuals in order to determine the correlated signal.  This is not useful because the resulting clock errors in the timing residuals for a particular pulsar will be modified by the fitting process that has been carried out for that specific pulsar.   In order to illustrate this effect, we simulate the timing residuals for three synthetic pulsars.  Each pulsar has the same ToA uncertainty (50\,ns) and is sampled every 14 days.  The first pulsar has continuous observations from the year 1994 to 2011.  The second pulsar only has observations until 2001 and the third pulsar has a gap of a few years around the year 2003.  The observations are simulated assuming that TT(BIPM11) is perfect, but residuals are formed using the TT(TAI) timescale. These resulting pre-fit residuals therefore exhibit the differences between TT(TAI) and TT(BIPM11) along with 50\,ns of additional white noise. A standard pulsar-timing fit is carried out for each pulsar (i.e., the pulse frequency and its first time derivative are fitted for).  For the third pulsar, we also include an arbitrary phase jump between the early and late observations.    The pre- and post-fit timing residuals are shown in left and right panels of Figure~\ref{fg:sim1} respectively.  The pre-fit residuals, in the left-hand panel, clearly are correlated and take the expected form shown in Figure~\ref{fg:tai_bipm} (note that due to the nature of the simulations these residuals are the inverse of the clock error shown in Figure~\ref{fg:tai_bipm}).  However, the fitting procedure significantly modifies the shape of the residuals and the post-fit residuals have a correlation coefficient significantly less than 1.

The average of our simulations is shown in the top panel of Figure~\ref{fg:compareWeight}.  The average was simply calculated as the mean residual for each sample\footnote{Note that previous work based on the average timing residuals, such as Rodin \& Chen (2011), use a Wiener filter when averaging their data sets. We are able to obtain the mean residual for each sample because our simulations have equal weighting and identical sampling.}.  Even though the average does show some of the features of the clock error, it does not model the clock errors perfectly because of the fitting of the pulsar timing models.  For instance, the weighted average also leads to a step-change around the year 2001; this occurs when there is a change in the number of pulsars contributing to the average. Such a procedure will therefore only work if all the pulsars have an equal data span and the same timing model fits are applied to all pulsars.  


$\Delta_c(t)$ values obtained using our method and their uncertainties are shown in the central panel of Figure~\ref{fg:compareWeight}.  Clearly this procedure successfully models the features resulting from the steering of TAI.  We note that the uncertainties on $\Delta_c(t)$ remain relatively small between the years 2001 and 2004 even though we have simulated data for only one pulsar during this time.  However, in reality, it is not possible to distinguish between clock errors and pulsar timing noise using data from a single pulsar.  To account for both timing noise and clock errors we first assume that all the noise in a given pulsar's timing residuals is timing noise.  We model the spectrum of this noise (using the \textsc{spectralModel} plugin to \textsc{tempo2}) and use a generalised least-squares-fitting procedure (Coles et al. 2011)\nocite{chc+11}  to account for the timing noise when fitting the pulsar timing model and the $\Delta_c(t)$ parameters.   The bottom panel in Figure~\ref{fg:compareWeight} demonstrates how the error bars on $\Delta_c(t)$ significantly increase when the noise for each pulsar is modelled as timing noise. If, as in this case, a significant clock error is measured, then these errors can be included in the timing procedure and the process iterated to determine true spectral models of the timing noise and hence the true error bars on $\Delta_c(t)$. 

\begin{figure}
\includegraphics[width=6cm,angle=-90]{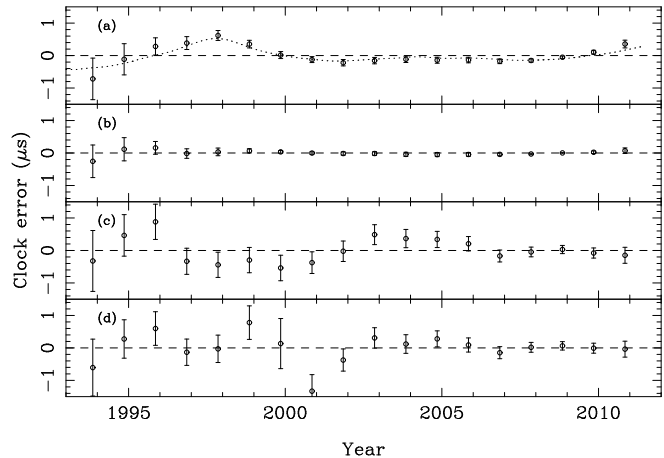}
\caption{Results obtained for all 19 PPTA pulsars, but with simulated arrival times.  In (a) only white noise and the clock error is simulated,  (b) only white noise and c) only uncorrelated red noise.  In (d) the same simulations are carried out as in (c), but PSR J0437$-$4715 is not included.}\label{fg:sim_results}
\end{figure}

We emphasise that the $\Delta_c(t)$ measurements and their uncertainties are not necessarily independent.   The effect of fitting, irregular sampling, differing data spans and the linear interpolation between adjacent grid points will all lead to correlated $\Delta_c(t)$ values.  The amount of correlation can be determined from the covariance matrix of the fit.


In order to test our algorithm with realistic data, we formed simulated data with the exact sampling and ToA uncertainties as in the observations of the 19 actual pulsars.  Initially we added no timing noise, but as before, simulated the data using TT(BIPM11) and formed timing residuals using TT(TAI).  In the top panel of Figure~\ref{fg:sim_results} we demonstrate that our algorithm correctly recovers the expected signal.  This demonstrates that the data sets and ToA uncertainties are such that our algorithm should correctly recover the irregularities in TT(TAI).

We wish to confirm that our method does not incorrectly lead to an error in TT if none exists. In Figure~\ref{fg:sim_results}b we show the results after forming the timing residuals using TT(BIPM11).  In this case, no clock error exists in the data (the timing residuals are purely white noise). We correctly find no significant $\Delta_c(t)$ values.  For Figure~\ref{fg:sim_results}c we have simulated white data and added uncorrelated red noise to represent timing noise. The resulting clock function does not show an unexpected large signal, but does show some correlated structure in the resulting data points.  As discussed in more detail below, this is mainly caused by the timing residuals for PSR~J0437$-$4715 dominating the fit.  In Figure~\ref{fg:sim_results}d we reproduce the analysis, but do not include PSR~J0437$-$4715.  The addition of timing noise therefore increases the error bars on $\Delta_c(t)$, but does not lead to an incorrect measurement of a clock error.

\section{Results and discussion}\label{sec:discussion}

\begin{figure}
\includegraphics[width=6cm,angle=-90]{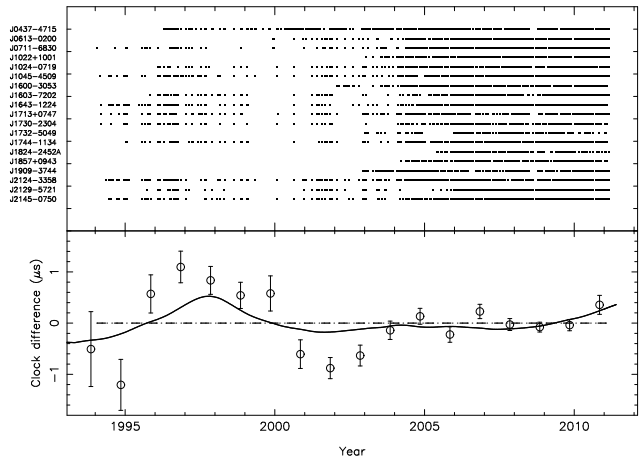}
\caption{The top panel shows the sampling for the 19 pulsars in our sample. The lower panel shows the difference between TT(BIPM11) and TT(TAI) as the solid line.  The data points indicate the difference between TT(PPTA11) and TT(TAI).}\label{fg:results}
\end{figure}

The top panel in Figure~\ref{fg:results} indicates the sampling for each of the 19 pulsars.  The solid line in the lower panel is the difference between TT(BIPM11) and TT(TAI).  This has been obtained by 1) sampling the expected signal at the same times as our measured clock function, 2) fitting a quadratic polynomial using an unweighted least-squares-fit and 3) removing this quadratic polynomial from the expected signal.  The data points in the Figure represent $\Delta_c(t)$, the difference between the pulsar-based timescale TT(PPTA11) and TT(TAI).   All the recent data are consistent within 2$\sigma$ with the expectation from TT(BIPM11), however some marginally significant differences are seen in the earlier data.  A careful statistical analysis is non-trivial as 1) the error bars on the data points are correlated and 2) the clock errors are constrained so that they do not include a quadratic polynomial.   The reduced-$\chi^2$ value obtained by comparing the expected clock signal with our data is 2.7.  This simple statistical test assumes that each data point is independent, but the value does indicate that there are no large discrepancies between the expected clock errors and the measurements.

The most obvious discrepancies between our values and the expectation occur between the years 1995 and 2003.  However, 1) our pulsar data set has sparse sampling around this time and has not been corrected for dispersion measure variations and 2) the observed discrepancies would require an error in the frequency of TT(BIPM11) of $\sim 10^{-14}$ whereas the uncertainty on this frequency is thought to be $\sim 1\times10^{-15}$ around the year 2003 (Petit 2003)\nocite{pet03b}.  This suggests that the discrepancies result from the determination of $\Delta_c(t)$.  There may be sufficient archival observations from other observatories to improve the clock error estimates during this period and thus to confirm or deny these possible errors in TT(BIPM11) and its estimated uncertainty.

\begin{figure}
\includegraphics[width=6cm,angle=-90]{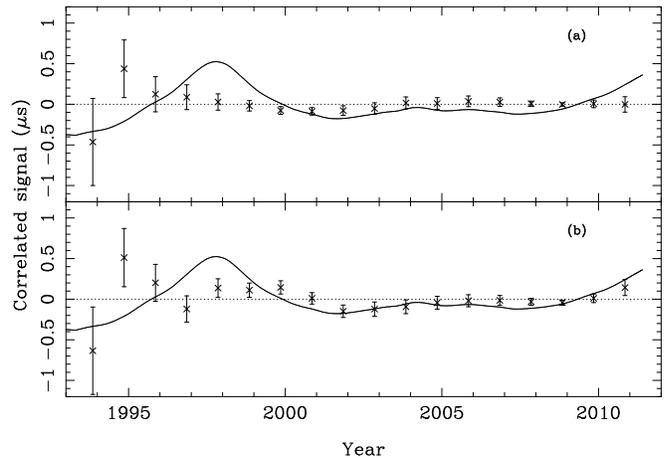}
\caption{The correlated signal caused by (a) errors in the Solar System ephemeris and (b) from one realisation of a gravitational wave background with dimensionless amplitude of $10^{-15}$. The solid line indicates the expected correlated signal caused by the steering of TAI.}\label{fg:ephem_gwb}
\end{figure}

It is possible that errors in the solar system ephemeris could lead to correlated signals in the timing residuals. To see the maximum size of any such signal in our data,  we have simulated observations using the same sampling and ToA uncertainties as the real data using the JPL DE421 solar system ephemeris, but without any clock errors. We then processed the data using the earlier JPL DE414 solar system ephemeris. The resulting estimate of the ``clock errors" are shown in the top panel of Figure~\ref{fg:ephem_gwb}. The maximum deviation for recent data is $<100$\,ns. As we use the most recent ephemeris, DE421, for our analysis it is likely that the actual correlated signal caused by the planetary ephemeris is significantly smaller than this.

In order to test whether a gravitational-wave background signal could be mis-identified as an error in the terrestrial time standard, we have simulated multiple realisations of a gravitational-wave background (Hobbs et al. 2009)\nocite{hjl+09} with a dimensionless strain amplitude of $10^{-15}$. This amplitude is typical of that expected for a background created by coalescing supermassive black-hole binary systems (e.g., Sesana, Vecchio \& Colacino 2008)\nocite{svc08}.  For each simulation we use the real sampling and ToA uncertainties as in the actual observations.  The results from our algorithm for one realisation are shown in the bottom panel of Figure~\ref{fg:ephem_gwb}. This shows that, for current data spans, it is unlikely that such a signal will significantly affect the stability of the pulsar timescale.  However, with increasing data lengths and with improvements in the ToA precision achievable, the gravitational-wave background could become a significant factor.  In this case the clock estimation algorithm would need to be modified to make it orthogonal to the gravitational wave background.  

From our data, we therefore conclude that
\begin{itemize}
\item{the difference between TT(TAI) and TT(BIPM11) can be detected using pulsar data and that this difference, as expected, results from the deliberate steering of TAI.}
\item{there are no large unexpected errors in TT(BIPM11) over our data span.}
\item{the variations in TT(TAI) are at a significant level compared with the precision of current pulsar timing array observations. We note that Guinot (1988)\nocite{gui88} and subsequent papers from the clock community have already pointed out that TT(TAI) is not suitable for high time precision pulsar experiments and that TT(BIPM) should always be used. We confirm that TT(BIPM11) is adequate for current millisecond pulsar timing experiments.} 
\end{itemize}

Our results do not show the steering of TAI as clearly as in Figure~\ref{fg:sim_results}a.   This is mainly because the timing residuals for PSR~J0437$-$4715 dominate the data set:  it is the only pulsar that was observed around the year 2000 and has a large number of observations and very small ToA uncertainties. However, the statistical properties of the timing residuals for this pulsar suddenly change around the year 2006.  Prior to this date, observations were made in the 20\,cm band and have not been corrected for dispersion measure variations.  After this date, the observations were made with new instrumentation, in the 10\,cm band and the dispersion measure variations have been measured and removed.  Unfortunately, we do not currently have any method that can model highly non-stationary noise.  The statistical model of the timing noise is applicable over the entire data set and, as such, is not optimal for any individual section.  Future algorithmic developments may allow the effects of non-stationary noise to be included in our standard analysis procedure.  Prior to 1996, our data sets are dominated by PSRs J1713$+$0747 and J1744$-$1134 which have significantly lower rms timing residuals than the other pulsars observed during that time. However, the effects of observations of just a few pulsars dominating the fitting procedures and the effects of non-stationary noise will, in subsequent work, be mitigated by including observations of more pulsars from other observatories.

The International Pulsar Timing Array (IPTA) project is a collaboration between three individual projects based in Europe, Australia and North America (Hobbs et al. 2010)\nocite{haa+10}.  The total number of pulsars being observed changes as new pulsars are discovered, but approximately 40 pulsars are currently being observed.  Of these, the ToA timing precisions for $\sim$30 should be better than $1 \mu$s.  Production of high-quality data sets is ongoing and will soon lead to a significantly improved pulsar timescale.  In the longer term, the Square Kilometre Array telescope (e.g., Cordes et al. 2004)\nocite{ckl+04} should be able to observe many hundreds of pulsars with a timing precision of 100\,ns or better.  If, as expected, pulsars are stable over long timescales at this level then such data sets should provide a long-term time standard that is competitive with the world's best terrestrial time standards.

\section{Conclusion}

We have developed a new algorithm for determining the correlated signal in the timing residuals for multiple pulsars.   Any errors in the reference timescale will lead to such a correlated signal.  By comparing our measurements of pulse arrival times to TT(TAI) we have confirmed that we can recover the effects of the deliberate steering of TAI. We have not identified any significant discrepancies with TT(BIPM11), but have noted a marginal discrepancy between 1995 and 2003.  Other phenomena, such as an isotropic, stochastic, gravitational wave background will also lead to a correlated signal in pulsar data sets, but we show that such phenomena are not likely to affect our results.

 In the future it is likely that pulsar data sets will be processed in a manner that will simultaneously identify: irregularities in the time standard; errors in the solar system ephemeris; and gravitational waves.  By combining observations from numerous telescopes, such future data sets will significantly improve on the pulsar-based timescale that is presented here.    
 
\section*{Acknowledgments} 

The Parkes radio telescope is part of the Australia Telescope which is funded by the Commonwealth of Australia for operation as a National Facility managed by CSIRO.  The PPTA project was initiated with support from RNM's Australian Research Council (ARC) Federation Fellowship (FF0348478). GH acknowledges support from the National Natural Science Foundation of China (NSFC) \#10803006 and \#11010250  and the Australian Research Council \#DP0878388.  JPWV was supported by the European Union under Marie-Curie Intra-European Fellowship 236394.  We thank the referee,  G{\'e}rard Petit, for extremely useful comments on the manuscript.

\bibliography{modrefs,psrrefs,crossrefs}

14 gid=286816
20 ctime=1344993879
20 atime=1344993900
24 SCHILY.dev=234881026
23 SCHILY.ino=12731850
18 SCHILY.nlink=1


\begin{thebibliography}{{{Hobbs}, {Edwards} \& {Manchester} }{2006}}

\bibitem[\protect\citename{{Arias}, {Panfilo} \& {Petit} }{2011}]{app11}
{Arias}~E.~F., {Panfilo}~G., {Petit}~G., 2011, Metrologia, 48, S145

\bibitem[\protect\citename{{Champion} {\rm et~al. }}{2010}]{chm+10}
{Champion}~D.~J. {\rm et~al.}, 2010, ApJ, 720, L201

\bibitem[\protect\citename{{Coles} {\rm et~al. }}{2011}]{chc+11}
{Coles}~W., {Hobbs}~G., {Champion}~D.~J., {Manchester}~R.~N.,
  {Verbiest}~J.~P.~W., 2011, MNRAS, 418, 561

\bibitem[\protect\citename{Cordes {\rm et~al. }}{2004}]{ckl+04}
Cordes~J.~M., Kramer~M., Lazio~T. J.~W., Stappers~B.~W., Backer~D.~C.,
  Johnston~S., 2004, New Astronomy Reviews, 48, 1413

\bibitem[\protect\citename{{Edwards}, {Hobbs} \& {Manchester} }{2006}]{ehm06}
{Edwards}~R.~T., {Hobbs}~G.~B., {Manchester}~R.~N., 2006, MNRAS, 372, 1549

\bibitem[\protect\citename{Foster \& Backer }{1990}]{fb90}
Foster~R.~S., Backer~D.~C., 1990, ApJ, 361, 300

\bibitem[\protect\citename{Guinot \& Petit }{1991}]{gp91}
Guinot~B., Petit~G., 1991, AA, 248, 292

\bibitem[\protect\citename{Guinot }{1988}]{gui88}
Guinot~B., 1988, AA, 192, 370

\bibitem[\protect\citename{Hellings \& Downs }{1983}]{hd83}
Hellings~R.~W., Downs~G.~S., 1983, ApJ, 265, L39

\bibitem[\protect\citename{{Hobbs}, {Edwards} \& {Manchester} }{2006}]{hem06}
{Hobbs}~G.~B., {Edwards}~R.~T., {Manchester}~R.~N., 2006, MNRAS, 369, 655

\bibitem[\protect\citename{{Hobbs} {\rm et~al. }}{2009}]{hjl+09}
{Hobbs}~G. {\rm et~al.}, 2009, MNRAS, 394, 1945

\bibitem[\protect\citename{{Hobbs} {\rm et~al. }}{2010}]{haa+10}
{Hobbs}~G., {Archibald}~A., {Arzoumanian}~Z., {Backer}~D., {Bailes}~M.,
  {Bhat}~N.~D.~R., {Burgay}~M., et~al., 2010, Classical and Quantum Gravity,
  27(8), 084013

\bibitem[\protect\citename{Irwin \& Fukushima }{1999}]{if99}
Irwin~A.~W., Fukushima~T., 1999, AA, 348, 642

\bibitem[\protect\citename{{Keith} \& {et al.} }{2012}]{kcs+12}
{Keith}~R.~N., {et al.}, 2012, MNRAS, Submitted

\bibitem[\protect\citename{{Lyne} {\rm et~al. }}{2010}]{lhk+10}
{Lyne}~A., {Hobbs}~G., {Kramer}~M., {Stairs}~I., {Stappers}~B., 2010, Science,
  329, 408

\bibitem[\protect\citename{{Manchester} \& {et al.} }{2012}]{mhb+12}
{Manchester}~R.~N., {et al.}, 2012, PASA, Submitted

\bibitem[\protect\citename{Petit \& Tavella }{1996}]{pt96}
Petit~G., Tavella~P., 1996, AA, 308, 290

\bibitem[\protect\citename{Petit }{2003}]{pet03b}
Petit~G., 2003, in 35th Annual Precise Time and Time Interval (PTTI) Meeting,
  San Diego, December 2003.
\newblock p.~307, TT(BIPM2003),
  http://tycho.usno.navy.mil/ptti/ptti2003/paper29.pdf

\bibitem[\protect\citename{{Rodin} \& {Chen} }{2011}]{rc11}
{Rodin}~A.~E., {Chen}~D., 2011, Astronomy Reports, 55, 622

\bibitem[\protect\citename{{Rodin} }{2008}]{rod08}
{Rodin}~A.~E., 2008, MNRAS, 387, 1583

\bibitem[\protect\citename{{Rodin}, {Kopeikin} \& {Ilyasov} }{1997}]{rki97}
{Rodin}~A.~E., {Kopeikin}~S.~M., {Ilyasov}~Y.~P., 1997, Acta Cosmologica, 23,
  163

\bibitem[\protect\citename{{Sesana}, {Vecchio} \& {Colacino} }{2008}]{svc08}
{Sesana}~A., {Vecchio}~A., {Colacino}~C.~N., 2008, MNRAS, 390, 192

\bibitem[\protect\citename{{Verbiest} {\rm et~al. }}{2008}]{vbv+08}
{Verbiest}~J.~P.~W. {\rm et~al.}, 2008, ApJ, 679, 675

\bibitem[\protect\citename{{Verbiest} {\rm et~al. }}{2009}]{vbc+09}
{Verbiest}~J.~P.~W. {\rm et~al.}, 2009, MNRAS, 400, 951

\bibitem[\protect\citename{{You} {\rm et~al. }}{2007}]{yhc+07}
{You}~X.~P. {\rm et~al.}, 2007, MNRAS, 378, 493

\end{thebibliography}
\bibliographystyle{mn}

\appendix
\section{Software instructions}

The algorithm described in this paper has been included in the \textsc{clock} plugin to \textsc{tempo2}.  Assuming that the user has a set of parameter and arrival time files (\verb|.par| and \verb|.tim|),  the following procedure can be followed:

\begin{itemize}
\item Obtain a spectral model for the timing noise in each pulsar and the corresponding covariance function: 
\begin{verbatim}
> tempo2 -gr spectralModel -f psr1.par psr1.tim
\end{verbatim}
(repeat for each pulsar).
\item Create a global parameter file (\verb|global.par|) that contains the required realisation of Terrestrial Time and a set of \verb|IFUNC| parameters that define the $t_i$ grid points described in the text:
\begin{verbatim}
# global.par
CLK TT(TAI)
EPHEM DE421
SIFUNC 2 2
IFUNC1 49300 0 0
IFUNC2 49600 0 0
....
IFUNC22 55600 0 0
\end{verbatim}
where the \verb|IFUNC| values range from before the earliest observation to after the latest observation.  Note that the \verb|SIFUNC 2 2| selects linear interpolation between the grid points (the first `2' on this line) and states that this parameter should be fitted globally between the pulsars (the second `2' on this line).
\item The global fit must be constrained not to include an offset, linear or quadratic component.  The following should be included in the parameter file for the first pulsar:
\begin{verbatim}
CONSTRAIN IFUNC
\end{verbatim}
\item The clock plugin can now be run.  Typical usage is:
\begin{verbatim}
> tempo2 -gr clock -fitfunc globalDCM -global 
  global.par -f psr1.par psr1.tim -f psr2.par 
  psr2.tim -f psr3.par psr3.tim ...
\end{verbatim}
\end{itemize}

\end{document}